\documentstyle[12pt,epsfig]{article}
\input{epsf}

\newcommand{\be}{\begin{eqnarray}}
\newcommand{\ee}{\end{eqnarray}}
\newcommand{\bea}{\begin{eqnarray}}
\newcommand{\eea}{\end{eqnarray}}

\newcommand{\beq}{\begin{equation}}
\newcommand{\eeq}{\end{equation}}
\newcommand{\nn}{\nonumber}

\def\fun#1#2{\lower3.6pt\vbox{\baselineskip0pt\lineskip.9pt
\ialign{$\mathsurround=0pt#1\hfil##\hfil$\crcr#2\crcr\sim\crcr}}}

\begin{document}

\title{ QUARK--DIQUARK SYSTEMATICS OF BARYONS: \\
SPECTRAL INTEGRAL EQUATIONS FOR SYSTEMS COMPOSED BY  LIGHT QUARKS
}

\author{A.V. ANISOVICH,
        V.V. ANISOVICH\footnote{anisovic@thd.pnpi.spb.ru}, M.A. MATVEEV,\\
        V.A. NIKONOV,
        A.V. SARANTSEV and T.O. VULFS}

 \date{}

\maketitle

\begin{abstract}
For baryons composed by the light quarks ($q=u,d$) we write spectral
integral equation using the notion of two diquarks: (i)
axial--vector state, $D^{1}_{1}$, with the spin $S_D=1$ and isospin
 $I_D=1$
 and (ii) scalar one,
 $D^{0}_{0}$, with the spin $S_D=0$ and isospin
 $I_D=0$.
We present spectral integral equations for the
$qD^{0}_{0}$ and $qD^{1}_{1}$ states taking into account
quark--diquark confinement interaction.
\end{abstract}


\section{Introduction}

In the present paper we continue to investigate baryons as quark--diquark
systems, $qD^{0}_{0}$ and $qD^{1}_{1}$. Here we consider  in detail the
systematization which was suggested in  \cite{qd,qd2}.

In \cite{qd2},   a  realistic
classification of baryons was supposed. The introduction of diquarks
allowed us to get  a considerable decrease of excited states as compared to
the quark model results of \cite{Izgur-3q,Gloz,Petry}, though exceeding
experimental data \cite{PDG}. An additional decrease of states may be due to
 the overlap of $I=1/2$ states with $S=1/2$.

The classification obtained in \cite{qd2}  results in the linearity of trajectories
in the $(J,M^2)$ and $(n,M^2)$ planes; the trajectories are strictly
ordered. A particular feature of the scheme is the prediction of the
number of overlapping poles. The observation of two-pole and
three-pole structures in the complex-$M$ planes of partial amplitudes
is a primary task in the verification of the scheme, while the next task
consists in a writing and solving the spectral integral equation for
quark--diquark systems with the aim to reconstruct the interaction.
Such an equation should be similar to that written and solved before
for the $q\bar q$ system \cite{si-qq}.

Addressing  the quark-diquark scheme, we suppose that
the excited baryons  do not prefer to be
three-body systems of spatially separated colored quarks.
Instead, similarly to mesons, they are  two-body systems of quark and
diquark:
$ q_\alpha D^{\alpha} =
q_\alpha[\varepsilon^{\alpha\beta\gamma}q_\beta q_\gamma ]
$,
where $\varepsilon^{\alpha\beta\gamma}$ is three-dimensional totally
 antisymmetrical tensor which works  in
 color space. Below, we omit color indices imposing  the
 symmetry anzatz for the spin--flavor--coordinate variables of wave
 functions.

It is an old idea that a $qq$-system inside the baryon can be
regarded as a specific object -- diquark. Thus, interactions with a
baryon can be considered as interactions with quark, $q$, and
two-quark system, $qq$: such a hypothesis was used in \cite{Y_5vva65}
for the description of the high-energy hadron--hadron collisions. In
\cite{ida,Y_5licht,ono}, baryons were described as quark--diquark
systems. In hard processes on nucleons (or nuclei), the coherent
$qq$ state (composite diquark) can be responsible for interactions
in the region of large Bjorken-$x$ values, at $x \sim 2/3$; deep
inelastic scatterings were considered in the framework of such an
approach in \cite{Y_5vva75,sch,fec,kawabe,fred}. More detailed
considerations of the diquarks and their applications to different
processes may be found in \cite{Y_5diquark1,Y_5goeke,Y_5diquark2}.

\begin{figure}[h]
\centerline{\epsfig{file=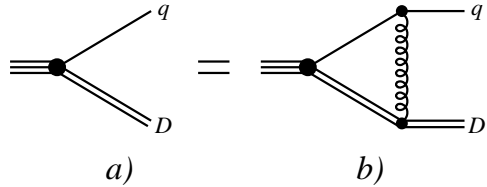,width=7.5cm}}
\caption{a,b)
Equation for quark--diquark system (the flavor-neutral interaction
denoted by helix-type line).
}
\label{Int}
\end{figure}

Here we concentrate our efforts on writing
 equations for pure  $qD^0_0$ and $qD^1_1$ systems; the contribution of
 three-quark states is neglected. Spectral integral equations for
 $qD^0_0$ and $qD^1_1$ systems are shown in Fig. \ref{Int}: the double
 line means diquark ($qD^0_0$ or $qD^1_1$), the flavor-neutral singular
interaction is denoted by helix-type line. The flavor-neutrality
 of the interaction results in the absence of mixing of  $qD^0_0$ and
 $qD^1_1$ states.

 The further presentation is organized as follows. In section 2,
 we give the elements of
 technique, which are needed for writing  spectral integral
 equations for $qD^0_0$ or $qD^1_1$ systems. In section 3, we discuss
 confinement singularities while in sections 4 and 5 we present the
 spectral integral equations.

\section{
 Technique for the description of fermions  with large spin  }

Recall some necessary properties of  angular momentum operators
for  two-particle systems and baryon projection operators. For more
detail, see \cite{book3} and references therein.

 \subsection{ Angular momentum operators for the
two-particle systems}

As in
\cite{book3,oper},
we use angular momentum operators
$X^{(L)}_{\mu_1\ldots\mu_L}(k^\perp)$,
$Z^{\alpha}_{\mu_1\ldots\mu_L}(k^\perp)$ and projection operator
$O^{\mu_1\ldots\mu_L}_{\nu_1\ldots\nu_L}(\perp P)$. Let us recall their
definition.

The operators are constructed from relative momenta
$k^\perp_\mu$ and tensor $g^\perp_{\mu\nu}$. Both are
orthogonal to the total momentum of the system:
\beq
k^\perp_\mu=\frac12
g^\perp_{\mu\nu}(k_1-k_2)_\nu =k_{1\nu} g^{\perp P}_{\nu\mu}
=-k_{2\nu} g^{\perp P}_{\nu\mu} ,  \qquad   g^{\perp }_{\mu\nu} \equiv
g^{\perp P}_{\mu\nu}=g_{\mu\nu}-\frac{P_\mu P_\nu}{s}\;.
\eeq
The operator for $L=0$ is a scalar (we write $X^{(0)}(k^\perp)=1$)
and the operator for $L=1$ is a vector, $X^{(1)}_\mu=k^\perp_\mu $.
The operators $X^{(L)}_{\mu_1\ldots\mu_L}$ for $L\ge 1$ can be
written in the form of a recurrence relation:
\bea
X^{(L)}_{\mu_1\ldots\mu_L}(k^\perp)&=&k^\perp_\alpha
Z^{\alpha}_{\mu_1\ldots\mu_L}(k^\perp)  ,
\nonumber\\
Z^{\alpha}_{\mu_1\ldots\mu_L}(k^\perp)&=&
\frac{2L-1}{L^2}\Big (
\sum^L_{i=1}X^{{(L-1)}}_{\mu_1\ldots\mu_{i-1}\mu_{i+1}\ldots\mu_L}(k^\perp)
g^\perp_{\mu_i\alpha}
\nonumber \\
 -\frac{2}{2L-1}  \sum^L_{i,\ell=1 \atop i<\ell}
&g^\perp_{\mu_i\mu_\ell}&
X^{{(L-1)}}_{\mu_1\ldots\mu_{i-1}\mu_{i+1}\ldots\mu_{\ell-1}\mu_{\ell+1}
\ldots\mu_L\alpha}(k^\perp) \Big ).
\label{Vz}
\eea
We have the convolution equality
$X^{(L)}_{\mu_1\ldots\mu_{L}}(k^\perp)k^\perp_{\mu_L}=k^2_\perp
X^{(L-1)}_{\mu_1\ldots\mu_{L-1}}(k^\perp)$,
with $k^2_\perp\equiv k^\perp_{\mu}k^\perp_{\mu}$, together with
 tracelessness property of $X^{(L)}_{\mu\mu\mu_3\ldots\mu_{L}}=0$.
On this basis,
one can write down the normalization condition for
the angular momentum  operators:
\beq
\int\frac{d\Omega}{4\pi}
X^{(L)}_{\mu_1\ldots\mu_{L}}(k^\perp)X^{(L')}_{\mu_1\ldots\mu_{L}}
(k^\perp)
 = \alpha_L k^{2L}_\perp \; ,\quad
\alpha_L\ =\ \prod^L_{l=1}\frac{2l-1}{l}  ,
\label{Valpha}
\eeq
where the integration is carried out over spherical variables: $\int d\Omega/(4\pi)=1$.

Iterating Eq. (\ref{Vz}), one obtains the
following expression for the operator $X^{(L)}_{\mu_1\ldots\mu_L}$
at $L\ge 1$:
\bea
\label{Vx-direct}
&&X^{(L)}_{\mu_1\ldots\mu_L}(k^\perp)=
\alpha_L \Big [
k^\perp_{\mu_1}k^\perp_{\mu_2}k^\perp_{\mu_3}k^\perp_{\mu_4}
\ldots k^\perp_{\mu_L}-\frac{k^2_\perp}{2L-1} \nn\\
&&\times\Big(
g^\perp_{\mu_1\mu_2}k^\perp_{\mu_3}k^\perp_{\mu_4}\ldots
k^\perp_{\mu_L}
+g^\perp_{\mu_1\mu_3}k^\perp_{\mu_2}k^\perp_{\mu_4}\ldots
k^\perp_{\mu_L} + \ldots \Big)+\frac{k^4_\perp}{(2L\!-\!1)(2L\!-\!3)}
\nn \\
&&\times\Big(
g^\perp_{\mu_1\mu_2}g^\perp_{\mu_3\mu_4}k^\perp_{\mu_5}
k^\perp_{\mu_6}\ldots k^\perp_{\mu_L}
+
g^\perp_{\mu_1\mu_2}g^\perp_{\mu_3\mu_5}k^\perp_{\mu_4}
k^\perp_{\mu_6}\ldots k^\perp_{\mu_L}+
\ldots\Big)+\ldots\Big ].
\eea
For the projection operators, one has:
\bea
&&\hspace{-6mm}O= 1 ,\qquad O^\mu_\nu (\perp P)=g_{\mu\nu}^\perp \, ,\nn \\
&&\hspace{-6mm}O^{\mu_1\mu_2}_{\nu_1\nu_2}(\perp P)=
\frac 12 \left (
g_{\mu_1\nu_1}^\perp  g_{\mu_2\nu_2}^\perp \!+\!
g_{\mu_1\nu_2}^\perp  g_{\mu_2\nu_1}^\perp  \!- \!\frac 23
g_{\mu_1\mu_2}^\perp  g_{\nu_1\nu_2}^\perp \right ).
\eea
For higher states, the operator can be calculated, using the
recurrent expression:
\bea
&&O^{\mu_1\ldots\mu_L}_{\nu_1\ldots\nu_L}(\perp P)=
\frac{1}{L^2} \Big (
\sum\limits_{i,\ell=1}^{L}g^\perp_{\mu_i\nu_\ell}
O^{\mu_1\ldots\mu_{i-1}\mu_{i+1}\ldots\mu_L}_{\nu_1\ldots
\nu_{\ell-1}\nu_{\ell+1}\ldots\nu_L}(\perp P)
\nn \\
&&- \frac{4}{(2L-1)(2L-3)}
\times
\sum\limits_{i<\ell\atop k<m}^{L}
g^\perp_{\mu_i\mu_\ell}g^\perp_{\nu_k\nu_m}
O^{\mu_1\ldots\mu_{i-1}\mu_{i+1}\ldots\mu_{\ell-1}\mu_{\ell+1}\ldots\mu_L}_
{\nu_1\ldots\nu_{k-1}\nu_{k+1}\ldots\nu_{m-1}\nu_{m+1}\ldots\nu_L}(\perp P)
\Big ).\qquad
\eea
The projection operators
 obey the
relations:
\bea
O^{\mu_1\ldots\mu_J}_{\nu_1\ldots\nu_L}(\perp P)
X^{(L)}_{\nu_1\ldots\nu_L}(k^\perp)&=&
X^{(L)}_{\mu_1\ldots\mu_L}(k^\perp)\, ,\nn \\
O^{\mu_1\ldots\mu_L}_{\nu_1\ldots\nu_L}(\perp P) k_{\nu_1} k_{\nu_2}
\ldots k_{\nu_L} &=& \frac
{1}{\alpha_L}X^{(L)}_{\mu_1\ldots\mu_L}(k^\perp) .
\eea
 Hence, the product of  two $X^L(k_\perp)$ operators results in the Legendre
polynomials as follows:
\beq
X^{(L)(k_\perp)}_{\mu_1\ldots\mu_L}(p^\perp) (-1)^J
O^{\mu_1\ldots\mu_L}_{\nu_1\ldots\nu_L}(\perp P)
X^{(L)}_{\nu_1\ldots\nu_L}(k^\perp)\!=\!\alpha_L
(\sqrt{-p_\perp^2}\sqrt{-k_\perp^2})^J P_L(z),
\eeq
where $z\equiv -(p^\perp k^\perp)/(
\sqrt{-p_\perp^2}\sqrt{-k_\perp^2})$.

\subsection{Baryon projection operators}

{\bf (i) Projection operators for particles with $J=1/2 $.}

It is convenient to use the baryon wave functions
 $u^{(a)}(p)$ and $\bar u^{(a)}(p)$, which are normalized as
 \beq
  \bar u^{(a)}(p) u^{(b)} (p)=\delta_{ab}
 \eeq
  and obey the completeness  condition
 \beq
 \sum_{a=1,2} u^{(a)}(p) \bar u^{(a)}(p)=\frac{m+\hat p}{2m}\,.
 \eeq
  For a  baryon with fixed polarization, one has to substitute:
\beq
\frac{m+\hat p}{2m}\to
\frac{m+\hat p}{2m}\left(1 + \gamma_5 \hat S\right),
\eeq
 with the following normalization for the polarization vector
 $ S^2=-1$ and the constraint $(pS)=0$.

{\bf (ii) Projection operators for particles with $J>1/2 $.}

The wave function of a particle with the  momentum $p$,
 mass $m$ and spin $ J=j+\frac 12$
  is given by a four-spinor  tensor
$\Psi_{\mu_1\ldots\mu_j}$. It satisfies the constraints
\beq
\label{5-8}
(\hat p-M_{J^P})\Psi_{\mu_1\ldots\mu_j}=0, \quad
p_{\mu_i}\Psi_{\mu_1\ldots\mu_j}=0,\quad
\gamma_{\mu_i}\Psi_{\mu_1\ldots\mu_j}=0,
\eeq
 and the symmetry properties
 \bea
  \label{5-9}
\Psi_{\mu_1\ldots\mu_i\ldots\mu_\ell\ldots\mu_j}&=&
\Psi_{\mu_1\ldots\mu_\ell\ldots\mu_i\ldots\mu_j}\;,
\nonumber \\
g_{\mu_i\mu_\ell}\Psi_{\mu_1\ldots\mu_i\ldots\mu_\ell\ldots\mu_j}&=&
g^{\perp p}
_{\mu_i\mu_\ell}\Psi_{\mu_1\ldots\mu_i\ldots\mu_\ell\ldots\mu_j}=0 .
\eea

The equations (\ref{5-8}), (\ref{5-9}) define the structure of
denominator of the fermion propagator (projection operator),
which can be written in the following form:
\beq \label{5-10}
\sum\limits_{a=1}^{2J+1}
 \Psi_{\mu_1\ldots\mu_j}^{(a)}\bar\Psi_{\mu_1\ldots\mu_j}^{(a)}\equiv
F^{\mu_1\ldots\mu_j}_{\nu_1\ldots \nu_j}(p)=(-1)^j
\frac{M_{J^P}+\hat p}{2M_{J^P}} R^{\mu_1\ldots\mu_j}_{\nu_1\ldots \nu_j}(\perp p)\ .
\eeq
The operator
$R^{\mu_1\ldots\mu_j}_{\nu_1\ldots \nu_j}(\perp p)$
describes  tensor structure of the propagator.
It is equal to 1 for the ($J=1/2$) particle and is proportional to
$g^{\perp p}_{\mu\nu}-\gamma^\perp_\mu\gamma^\perp_\nu/3$ for the
particle with spin $J=3/2$ (recall that
$\gamma^\perp_\mu=g^{\perp p}_{\mu\nu}\gamma_\nu$).

The conditions (\ref{5-9}) are the same for fermion and boson
projection operators,  therefore  fermion projection operator
can be written as follows:
\beq
  \label{5-11}
R^{\mu_1\ldots\mu_j}_{\nu_1\ldots \nu_j}(\perp p)=
O^{\mu_1\ldots\mu_j}_{\alpha_1\ldots \alpha_j} (\perp p)
T^{\alpha_1\ldots\alpha_j}_{\beta_1\ldots \beta_j}(\perp p)
O^{\beta_1\ldots \beta_j}_{\nu_1\ldots\nu_j} (\perp p)\ .
\eeq
 The operator
  $T^{\alpha_1\ldots\alpha_j}_{\beta_1\ldots \beta_j}(\perp p)$
  can be expressed in a  simple form, so all
symmetry and orthogonality conditions are imposed by $O$-operators.
First, the $T$-operator is constructed of metric tensors only, which
act in the space of $\perp p$ and $\gamma^\perp$ matrices. Second, a
construction like
$ \gamma^\perp_{\alpha_i}\gamma^\perp_{\alpha_\ell}=
\frac12 g^\perp_{\alpha_i\alpha_\ell}+\sigma^\perp_{\alpha_i\alpha_\ell}$,
with
$\sigma^\perp_{\alpha_i\alpha_\ell}=\frac 12
(\gamma^\perp_{\alpha_i}\gamma^\perp_{\alpha_\ell}-
\gamma^\perp_{\alpha_\ell}\gamma^\perp_{\alpha_i}$),
gives zero, when being multiplied by the
$O^{\mu_1\ldots\mu_j}_{\alpha_1\ldots \alpha_j}$-operator:
the first term is due to the tracelessness
condition and the second one to symmetry properties. The only
structures, which can be thus constructed, are
$g^\perp_{\alpha_i\beta_\ell}$ and $\sigma^\perp_{\alpha_i\beta_\ell}$.
Moreover, taking into account the symmetry properties of the
$O$-operators,  one can use any pair of indices from sets
$\alpha_1\ldots\alpha_j$ and $\beta_1\ldots \beta_j$, for example,
$\alpha_i\to \alpha_1$ and $\beta_\ell\to \beta_1$. Then,
\beq
T^{\alpha_1\ldots\alpha_j}_{\beta_1\ldots \beta_j}(\perp p)=
\frac{j+1}{2j\!+\!1} \big( g^\perp_{\alpha_1\beta_1}-
\frac{j}{j\!+\!1}\sigma^\perp_{\alpha_1\beta_1} \big)
\prod\limits_{i=2}^{j}g^\perp_{\alpha_i\beta_i} .
\label{5-t1}
\eeq
 Since $R^{\mu_1\ldots\mu_n}_{\nu_1\ldots \nu_n}(\perp p)$ is
determined by convolutions of $O$-operators, see Eq. (\ref{5-11}),
we can replace in (\ref{5-t1})
\beq \label{5-13}
T^{\alpha_1\ldots\alpha_j}_{\beta_1\ldots \beta_j}(\perp p) \to
T^{\alpha_1\ldots\alpha_j}_{\beta_1\ldots \beta_j}( p) =
\frac{j+1}{2j\!+\!1} \big( g_{\alpha_1\beta_1}-
\frac{j}{j\!+\!1}\sigma_{\alpha_1\beta_1} \big)
\prod\limits_{i=2}^{j}g_{\alpha_i\beta_i} .
\eeq
 The coefficients in (\ref{5-13}) are chosen to satisfy the constraints
(\ref{5-8}) and  convolution condition:
\beq  \label{5-14}
F^{\mu_1\ldots\mu_j}_{\alpha_1\ldots \alpha_j}( p) F^{\alpha_1\ldots
\alpha_j}_{\nu_1\ldots \nu_j}( p)=(-1)^j
F^{\mu_1\ldots\mu_j}_{\nu_1\ldots \nu_j}( p) \;.
\eeq
Spin projection operators are considered in a more detail
in \cite{book3}.

\section{Confinement singularities}

Confinement singularities were applied to the calculation of levels in
meson sector, to $q\bar q$ systems \cite{si-qq} and heavy quark ones
$Q\bar Q$ \cite{si-bb,si-cc}. Here we apply confinement singularities
to the quark--diquark compound states.

{\bf (i) Meson sector}

The linearity of the $q\bar q$-trajectories in the ($n,M^2$) planes in
meson sector \cite{syst} (experimentally,  up to large $n$ values,
 $n\le 7$) provides us the $t$-channel singularity $V_{conf}\sim 1/q^4$
which creates the barrier
$V_{conf}\sim r$ in the coordinate representation. The confinement
interaction is two-component \cite{si-qq}:
\bea \label{I1} &&V_{conf}=(I\otimes I)\,b_S\,r +
(\gamma_\mu\otimes \gamma_\mu)\,b_V\,r\ , \nn \\ &&b_S\simeq -b_V
\simeq 0.15\,\, {\rm GeV}^{-2}\ .
\eea
The position of $q\bar q$ levels and data on radiative decays tell us
that singular $t$-channel exchanges are necessary both in the scalar
($I\otimes I$) and vector ($\gamma_\mu\otimes \gamma_\mu$) channels.
The $t$-channel exchange interactions (\ref{I1}) can take place both
for white and color states, ${\bf c}={\bf 1}+{\bf 8}$, though, of
course, the color-octet interaction plays a dominant role.

The spectral integral equation for the meson-$q\bar q$ vertex (or
for  $q\bar q$ wave function of the meson)
was solved by introducing a cut-off into the interaction (\ref{I1}):
$r\to re^{-\mu r}$. The cut-off parameter is
small: $\mu\sim 1-10$ MeV; if $\mu$ is changing in this interval,
the $q\bar q$-levels  with $n\le 7$ remain practically the same.

In \cite{si-qq,si-bb,si-cc}, the spectral integral equations were
solved in the momentum representation -- this is natural, since we used
dispersion integration technique (see discussion in
\cite{book3}). In this representation, the interaction is re-written
as follows:
  \bea
  re^{-\mu r}\to
  8\pi\left(\frac{4\mu^2}{(\mu^2-q^2_\perp)^3}-\frac{1}{(\mu^2-q^2_\perp)^2}
  \right).
  \label{1-1-2}
  \eea

  {\bf (ii) Quark--diquark sector}

Bearing in mind that in the framework of spectral
integration (as in dispersion technique) the total energy is not
conserved, we have to write
  \bea
  q^2_\perp=(k^\perp_1-k^{'\perp}_1)_\mu(-k^\perp_2+k^{'\perp}_2)_\mu\
  \label{1-1-3}
  \eea
for the momentum transfer, where $k_1$ and $k_2$ are the momenta
of the initial quark and diquark, while $k'_1$ and $k'_2$ are
those after the interaction. The index $\perp$ means that we use
components perpendicular to total momentum $p$ for the initial
state and to $p'$ for the final state:
\bea
&&k^\perp_{i\,\mu}=g^{\perp p}_{\mu\nu}k_{i\,\nu}\ , \quad
g^{\perp p}_{\mu\nu}=g_{\mu\nu}-\frac{p_\mu p_\nu}{p^2}\ , \quad
p=k_1+k_2 \ , \quad p^2=s\, ,
\nn \\
&& k'^{\perp}_{i\,\mu}=g^{\perp p'}_{\mu\nu}k'_{i\,\nu}\ , \quad
g^{\perp p'}_{\mu\nu}=g_{\mu\nu}-\frac{p'_\mu p'_\nu}{p^{'2}}\ ,
\quad p'=k'_1+k'_2\, , \quad p'^2=s'  .
\label{1-1-4}
\eea
Generally, we can write for the $t$-channel interaction block:
  \bea
  \label{1-1-5}
  I_N(q^2_\perp)&=&\frac{4\pi(N+1)!}{(\mu^2-q^2_\perp)^{N+2}}
  \sum^{N+1}_{n=0}\left(\mu+\sqrt{q^2_\perp}\right)^{N+1-n}
  \left(\mu-\sqrt{q^2_\perp}\right)^{n}
  \eea

\section{The $qD^0_0$  systems  }

First, let us present vertices for the transitions $N_{J^P}\to qD^0_0$
and the block of the confinement interaction (transition
$qD^0_0\to qD^0_0$). Convoluting them, see Fig. \ref{Int},  we obtain
spectral integral equation for the $qD^0_0$  system. This equation is a
modified Bethe--Salpeter equation \cite{WSalpeter} (see also
\cite{RHersbach,Rgross,Rmaung}) re-written in terms of the dispersion
integrals \cite{WMandelstam}.

\subsection {Vertices for the transition $N_{J^P}\to qD^0_0$}

{\bf (i) The $N_{J^+}$ vertices with even $L$}

The vertex for $J=L+1/2$ with even $L$ is equal to:
\bea \label{A-22}
&&\Phi^+(k_2)\bar u(k_1)X^{(L)}_{\mu_1\ldots\mu_L}(k^\perp)
G^{(L,J^+)}(k^2_\perp)\psi_{(J^+)\mu_1\ldots\mu_j}(p)
\nn \\
&&\equiv
\Phi^+(k_2)\bar u(k_1)S_{\mu_1\ldots\mu_j}^{(L,J^+)}(k_1,k_2)G^{(L,J^+)}(k^2_\perp)
\psi_{(J^+)\mu_1\ldots\mu_j}(p),
\quad j=L, \quad
\eea
while for $J=L-1/2$ it is
\bea  \label{A-23}
&&\Phi^+(k_2)\bar u (k_1)i\gamma_5\gamma_\nu X^{(L)}_{\mu_1\ldots\mu_{L-1}\nu}(k^\perp)
G^{(J^+)}(k^2_\perp)
\psi_{(J^+)\mu_1\ldots\mu_j}(p)
\nn \\
&&\equiv
\Phi^+(k_2)\bar u(k_1)S_{\mu_1\ldots\mu_j}^{(L,J^+)}(k_1,k_2)G^{(L,J^+)}(k^2_\perp)
\psi_{(J^+)\mu_1\ldots\mu_j}(p),
\quad j=L-1\, .\quad
\eea

{\bf (ii) The $N_{J^-}$ vertices with odd $L$}

For odd $L$, the vertex for $J=L+1/2$ is equal to:
\bea \label{A-24}
&&\Phi^+(k_2)\bar u (k_1)X^{(L)}_{\mu_1\ldots\mu_L}(k^\perp)
G^{(J^-)}(k^2_\perp)\psi_{(J^-)\mu_1\ldots\mu_j}(p)
\nn \\
&&\equiv
\Phi^+(k_2)\bar u(k_1)S_{\mu_1\ldots\mu_j}^{(L,J^-)}(k_1,k_2)G^{(L,J^-)}(k^2_\perp)
\psi_{(J^+)\mu_1\ldots\mu_j}(p),
\quad j=L, \quad
\eea
while for $J=L-1/2$:
\bea  \label{A-25}
&&\Phi^+(k_2)\bar u (k_1)i\gamma_5\gamma_\nu X^{(L)}_{\mu_1\ldots\mu_{L-1}\nu}(k^\perp)
G^{(J^-)}(k^2_\perp)
\psi_{(J^-)\mu_1\ldots\mu_j}(p)
\nn \\
&&\equiv
\Phi^+(k_2)\bar u(k_1)S_{\mu_1\ldots\mu_j}^{(L,J^-)}(k_1,k_2)G^{(L,J^-)}(k^2_\perp)
\psi_{(J^+)\mu_1\ldots\mu_j}(p),
\quad j=L-1, \quad
\eea
We denote
vertices (\ref{A-22})-(\ref{A-25}) as follows:
\bea  \label{A-26}
&&
G^{(L,J^P)}_{qD^0_0}(k_1,k_2;p)=
\Phi^+(k_2)\bar u (k_1)G^{(L,J^P)}_{(qD^0_0)\mu_1\ldots\mu_j}(k_1,k_2)
\psi_{\mu_1\ldots\mu_j}^{(J^P)}(p)
\nn \\
&&\equiv
\Phi^+(k_2)\bar u (k_1)S^{(L,J^P)}_{(qD^0_0)\mu_1\ldots\mu_j}(k_1,k_2)
G^{(L,J^P)}_{qD^0_0}(k^2_\perp)
\psi_{\mu_1\ldots\mu_j}^{(J^P)}(p),
\quad j=J-\frac 12 \, .\quad
\eea
The wave function of the $qD^0_0$ system reads:
\bea              \label{A-26wf}
&&\psi_{(qD^0_0)\mu_1\ldots\mu_j}^{(L,J^P)}(k_1,k_2)=
S^{(L,J^P)}_{(qD^0_0)\mu_1\ldots\mu_j}(k_1,k_2)
\frac{G^{(L,J^P)}_{qD^0_0}(k^2_\perp)}{s-M^2_{J^P}}\nn\\
&&\equiv
S^{(L,J^P)}_{(qD^0_0)\mu_1\ldots\mu_j}(k_1,k_2)\psi^{(L,J^P)}_{qD^0_0}(s),
\eea
where $\psi^{(L,J^P)}_{qD^0_0}(s)$ is its invariant part.

\subsection{Confinement singularities in $qD^0_0$ interaction amplitude}

Introducing the momenta of quarks and diquarks, we must remember that
total energies are not conserved in the spectral integrals (like
in dispersion relations). Hence, in general case $s\neq s'$.

The interaction amplitude in the $qD^0_0$ system is written  in the following form:
\bea \label{1-1-6}
&&{\rm S}:\quad
\sum\limits_N \alpha^{(N)}_S(q^2_\perp)\Big(\bar u(k_1)\,I\, u(k'_1)\Big)\,
I^{(\mu\to 0)}_N(q^2_\perp) \Big(\Phi^+(k_2)\Phi(k'_2)\Big),
\nn \\
&&{\rm V}:\quad -\sum\limits_N \alpha^{(N)}_V(q^2_\perp)
\Big(\bar u(k_1)\,\gamma_\nu\,
u(k'_1)\Big)\, I^{(\mu\to 0)}_N(q^2_\perp)
\Big(\Phi^+(k_2)\,(k_2+k'_2)_\nu\,\Phi(k'_2)\Big).
\nn \\
\eea
 Recall, the singular block, $I^{(\mu\to 0)}_N(q^2_\perp)$, is
 given in
(\ref{1-1-5}), and the operator $\Phi(k_2)$ refers to  scalar diquarks.

Diquarks should be considered as composite particles, hence one can
expect an appearance of form factors in the interaction block,
correspondingly, $\alpha^{(N)}_S(q^2_\perp)$ and
$\alpha^{(N)}_V(q^2_\perp)$.

The sum of interaction terms in (\ref{1-1-6}) can be re-written in a
compact form:
\beq \label{1-1-7} \Phi^+(k_2)\bar u(k_1)\,
V_{D^0_0}^q(k_1,k_2;k'_1,k'_2)\, u(k'_1)\Phi(k'_2)
\eeq

\subsection{Spectral integral equation for $qD^0_0$  system   }

The spectral integral equation for $qD^0_0$  system reads (see Fig. \ref{Int}):
\bea  \label{A-34}
&&\Phi^+(k_2)\bar u (k_1)G^{(L,J^P)}_{(qD^0_0)\mu_1\ldots\mu_j}(k_1,k_2)
\psi_{\mu_1\ldots\mu_j}^{(J^P)}(p)
\nn \\
=&&\Phi^+(k_2)\bar u (k_1)\int\limits^\infty_{(m+M_{D^0_0})^2}
\frac{ds'}{\pi}\frac{d\phi_2(P';k'_1,k'_2)}{s'-M^2_{L,J^P}}
 V_{D^0_0}^q(k_1,k_2;k'_1,k'_2)\frac{\hat k'_1 +m}{2m}
 \nn \\
&&\qquad\qquad\qquad\qquad\qquad\qquad
\times G^{(L,J^P)}_{(qD^0_0)\mu'_1\ldots\mu'_j}(k'_1,k'_2)
\psi_{\mu'_1\ldots\mu'_j}^{(J^P)}(p)\ .
\eea
Here, the interaction block (the right-hand side of (\ref{A-34}))
is presented  using the spectral (dispersion relation) integral over
$ds'$, and $d\phi_2(P';k'_1,k'_2)$ is a standard phase-space
integral for the $qD^0_0$  system in the intermediate state.

It is suitable to work with equation re-written in the following
way:\\
{\bf (i)} the left-hand and right-hand sides of Eq. (\ref{A-34}) are
 convoluted with the spin operator of vertex (\ref{A-26}),\\
{\bf (ii)} the convoluted terms are integrated over final-state
phase space of the $qD^0_0$  system:
\beq
\label{A-34ps} d\phi_2(p;k_1,k_2)=\frac 12 (2\pi)^4
\delta^{(4)}(p-k_1-k_2) \prod\limits_{a=1,2}
\frac{d^3k_a}{(2\pi)^32k_{a0}} \, . \eeq
We obtain:
\bea  \label{A-35}
&&\int d\phi_2(p;k_1,k_2)
Sp\Big[S^{(L,J^P)}_{(qD^0_0)\nu_1\ldots\nu_j}(k_1,k_2)
\frac{\hat k_1+m}{2m}G^{(L,J^P)}_{(qD^0_0)\mu_1\ldots\mu_j}(k_1,k_2)
F_{\mu_1\ldots\mu_j}^{\nu_1\ldots\nu_j}(p)\Big] \nn\\
&&=\int d\phi_2(p;k_1,k_2)Sp\Big[
S^{(L,J^P)}_{(qD^0_0)\nu_1\ldots\nu_j}(k_1,k_2)\frac{\hat k_1+m}{2m}
\int\limits^\infty_{(m+M_{D^0_0})^2}
\frac{ds'}{\pi}\frac{d\phi_2(P';k'_1,k'_2)}{s'-M^2_{L,J^P}} \nn\\
&&\times V_{D^0_0}^q(k_1,k_2;k'_1,k'_2)\frac{\hat k'_1 +m}{2m}
G^{(L,J^P)}_{(qD^0_0)\mu'_1\ldots\mu'_j}(k'_1,k'_2)
F_{\mu'_1\ldots\mu'_j}^{\nu_1\ldots\nu_j}(p)
\Big].
\eea
The operators $F_{\mu_1\ldots\mu_j}^{\nu_1\ldots\nu_j}(p)$
and $S^{(L,J^P)}_{(qD^0_0)\mu_1\ldots\mu_j}(k_1,k_2)$ are given
 by Eqs. (\ref{5-10}) and (\ref{A-26}), correspondingly.
We can re-write (\ref{A-35}) incorporating the wave function
 $\psi^{(L,J^P)}_{qD^0_0}(s)$, which is determined in Eq.
 (\ref{A-26wf}):
\bea  \label{A-36}
&&\int d\phi_2(p;k_1,k_2)
Sp\Big[S^{(L,J^P)}_{(qD^0_0)\nu_1\ldots\nu_j}(k_1,k_2)
\frac{\hat k_1+m}{2m}S^{(L,J^P)}_{(qD^0_0)\mu_1\ldots\mu_j}(k_1,k_2)
F_{\mu_1\ldots\mu_j}^{\nu_1\ldots\nu_j}(p)\Big]
\nn \\
&&\times (s-M^2_{J^P})\psi^{(L,J^P)}_{(qD^0_0)}(s)
\nn\\
=&&\int d\phi_2(p;k_1,k_2)Sp\Big[
S^{(L,J^P)}_{(qD^0_0)\nu_1\ldots\nu_j}(k_1,k_2)\frac{\hat k_1+m}{2m}
\int\limits^\infty_{(m+M_{D^0_0})^2}
\frac{ds'}{\pi}d\phi_2(P';k'_1,k'_2) \nn\\
&&\times V_{D^0_0}^q(k_1,k_2;k'_1,k'_2)\frac{\hat k'_1 +m}{2m}
S^{(L,J^P)}_{(qD^0_0)\mu'_1\ldots\mu'_j}(k'_1,k'_2)\psi^{(L,J^P)}_{(qD^0_0)}(s')
F_{\mu'_1\ldots\mu'_j}^{\nu_1\ldots\nu_j}(p)
\Big] .
\eea

\section{Spectral integral equations for $qD^1_1$  systems
 with $I=3/2$ }

Here, we present equations for vertices, or wave functions, for the
transitions
 $\Delta_{J^P}\to qD^1_1$ with $P=\pm$. Below, for the shake of
 simplicity, we consider $\Delta_{J^P}^{++}$ state: in this case it is
 necessary to take into account one quark--diquark channel only,
 namely, $uD^{1,I_3=3/2}_1$. We omit
 isotopic indices, denoting the quark--diquark state as $qD^1_1$.

\subsection {Vertices for the transition $\Delta_{J^P}\to qD^1_1$}

The $qD^1_1$ states are characterized by the total spin of the quark and
diquark ($S=\frac 12 ,\frac 32$), orbital  momentum ($L$) and
total angular momentum ($J$). The parity ($P$) is determined by $L$.

The systematization performed in \cite{si-qq} favors, in the first
approximation, the  consideration of quantum numbers $S$ and $L$ as
good ones. Below, we follow this result.

Outgoing quark--diquark states with fixed $S$ read:
\bea \label{aDA-21}
S=\frac 12&:&\quad  \Phi_\nu^+ (k_2)\bar u(k_1)   i\gamma_5\gamma_\nu
\frac{\hat P +\sqrt s}{\sqrt s}\equiv\Phi_\nu^+ (k_2)\bar u(k_1)S^{1/2}_\nu (P)\ ,
\nn \\
S=\frac 32&:&\quad
 \Phi_\nu^+ (k_2)\bar u(k_1)
 \big(-\frac{2}{3} g^{\perp P}_{\nu\mu}+
\frac{1}{3}\sigma^{\perp P}_{\nu\mu} \big)
\frac{\hat P +\sqrt s}{\sqrt s}
\equiv\Phi_\nu^+ (k_2)\bar u(k_1)S^{3/2}_{\nu\mu}(P)\ .
\qquad
\eea
Here, we use the operator (\ref{1-1-6}) for spins
$1/2$ and $3/2$, substituting $M_{J^P}^2\to s$.

{\bf (i) The transition vertices $\Delta_{J^\pm}$  at $S=\frac 12 $}

The transition vertices for $J=L\pm 1/2$ read:
\bea \label{aDA-23}
&&J=L+ \frac 12\,:\quad\Phi^+_\nu(k_2)\bar u(k_1)S^{1/2}_\nu (P)
X^{(L)}_{\mu_1\ldots\mu_L}
(k^\perp)G^{(S,L,J^\pm)}(k^2_\perp) \psi^{(J^\pm)}_{\mu_1\ldots\mu_j}(p)
\nn \\
&&\equiv \Phi_\nu^+(k_2)\bar
u(k_1)S_{\nu\mu_1\ldots\mu_j}^{(1/2,L,J^\pm)}(k_1,k_2)
G^{(S,L,J^\pm)}(k^2_\perp) \psi^{(J^\pm)}_{\mu_1\ldots\mu_j}(p),
\quad j=L\ ,
\nn \\
&&J=L-\frac 12\,:\quad\Phi_\nu^+(k_2)\bar u (k_1)S^{1/2}_\nu (P)\, i\gamma_5\gamma_{\nu'}
 X^{(L)}_{\mu_1\ldots\mu_{L-1}\nu'}(k^\perp)
G^{(S,L,J^\pm)}(k^2_\perp)
\psi^{(J^\pm)}_{\mu_1\ldots\mu_j}(p)\nn \\
&&\equiv
\Phi_\nu^+(k_2)\bar u(k_1)S_{\nu\mu_1\ldots\mu_j}^{(1/2,L,J^\pm)}(k_1,k_2)
G^{(S,L,J^\pm)}(k^2_\perp)
\psi^{(J^\pm)}_{\mu_1\ldots\mu_j}(p),
\quad j=L-1\ .
\nn \\
\eea
Note that in (\ref{aDA-23}), as in (\ref{A-23}) and
(\ref{A-25}), we use  the axial--vector
 operator $i\gamma_5\gamma_{\nu'}$ for the decreasing rank of the
 vertex at fixed $L$; recall also that $P=+$ corresponds to even $L$
 and $P=-$ to odd ones.

{\bf (ii) The transition vertices $\Delta_{J^\pm}$  at $S=\frac 32 $}

The transition vertices for $J=|{\bf L}+{\bf \frac32}|$ and $P=(-1)^L$
are written as follows:
\bea \label{aDA-24}
&&J=L+\frac 32 :\quad\Phi_\nu^+ (k_2)\bar u(k_1)S^{3/2}_{\nu\mu_{L+1}}(P)
X^{(L)}_{\mu_1\ldots\mu_L}(k^\perp)G^{(3/2,L,J^\pm)}(k^2_\perp)
\psi^{(J^\pm)}_{\mu_1\ldots\mu_j}(p)
\nn \\
&\, &\equiv
\Phi_\nu^+(k_2)\bar u(k_1)S_{\nu\mu_1\ldots\mu_j}^{(3/2,L,J^\pm)}(k_1,k_2)
G^{(3/2,L,J^\pm)}(k^2_\perp)
\psi^{(J^\pm)}_{\mu_1\ldots\mu_j}(p),
\quad j=L+1\ ,
\nn \\
&&J=L+\frac 12 :\quad\Phi_\nu^+ (k_2)\bar u(k_1)S^{3/2}_{\nu\mu_{L}}(P)
i\gamma_5\gamma_{\nu_1}
X^{(L)}_{\nu_1\mu_1\ldots\mu_{L-1}}(k^\perp)G^{(3/2,L,J^\pm)}(k^2_\perp)
\psi^{(J^\pm)}_{\mu_1\ldots\mu_j}(p)
\nn \\
&\, &\equiv
\Phi_\nu^+(k_2)\bar u(k_1)S_{\nu\mu_1\ldots\mu_j}^{(3/2,L,J^\pm)}(k_1,k_2)
G^{(3/2,L,J^\pm)}(k^2_\perp)
\psi^{(J^\pm)}_{\mu_1\ldots\mu_j}(p),
\quad j=L\ ,
\nn \\
&&J=L-\frac 12 :\quad\Phi_\nu^+ (k_2)\bar u(k_1)S^{3/2}_{\nu\mu_{L-1}}(P)
\prod\limits_{a=1}^2(i\gamma_5\gamma_{\nu_a})
X^{(L)}_{\nu_1\nu_2\mu_1\ldots\mu_{L-2}}(k^\perp)
\nn \\
&&\qquad\qquad\qquad\times G^{(3/2,L,J^\pm)}(k^2_\perp)
\psi^{(J^\pm)}_{\mu_1\ldots\mu_j}(p)
\nn \\
&\, &\equiv
\Phi_\nu^+(k_2)\bar u(k_1)S_{\nu\mu_1\ldots\mu_j}^{(3/2,L,J^\pm)}(k_1,k_2)
G^{(3/2,L,J^\pm)}(k^2_\perp)
\psi^{(J^\pm)}_{\mu_1\ldots\mu_j}(p),
\quad j=L-1\ ,
\nn
\eea
\bea
&&J=L-\frac 32 :\quad\Phi_\nu^+ (k_2)\bar u(k_1)S^{3/2}_{\nu\mu_{L-2}}(P)
\prod\limits_{a=1}^3(i\gamma_5\gamma_{\nu_a})
X^{(L)}_{\nu_1\nu_2\nu_3\mu_1\ldots\mu_{L-3}}(k^\perp)
\nn \\
&&\qquad\qquad\qquad\times G^{(3/2,L,J^\pm)}(k^2_\perp)
\psi^{(J^\pm)}_{\mu_1\ldots\mu_j}(p)
\nn \\
&\, &\equiv
\Phi_\nu^+(k_2)\bar u(k_1)S_{\nu\mu_1\ldots\mu_j}^{(3/2,L,J^\pm)}(k_1,k_2)
G^{(3/2,L,J^\pm)}(k^2_\perp)
\psi^{(J^\pm)}_{\mu_1\ldots\mu_j}(p),
\quad j=L-2\ .\qquad
\eea

{\bf (iii) The invariant wave functions of the $qD^1_1$ systems}

The invariant wave functions of the $qD^1_1$ systems are determined by vertices
$G^{(S,L,J^\pm)}(k^2_\perp)$ as follows:
\beq              \label{aDA-26wf}
\frac{ G^{(S,L,J^\pm)}(k^2_\perp)}{s-M^2_{S,L,J^P}}
=
\psi^{(S,L,J^P)}_{qD^1_1}(s).
\eeq
Recall that relative momentum squared $k^2_\perp$ depends on $s$
only.

\subsection{Confinement singularities in $qD^1_1$ interaction block}

Introducing the momenta of quarks and diquarks, we should remember that
 total energies are not conserved in the spectral integrals (just as
in dispersion relations), so $s\neq s'$.

The interaction amplitude for $qD^1_1$ system is written  as
follows.\\
S-exchange:
\bea \label{aD1-1-6}
&&\Phi^+_\nu(k_2)\bar u(k_1)\, V_{qD^1_1 ;\nu\nu'}^{(S-ex)}(k_1,k_2;k'_1,k'_2)\,
 u(k'_1)\Phi_{\nu'}(k'_2)
 \nn\\
&=&
\sum\limits_N \beta^{(N)}_S(q^2_\perp)\Big(\bar u(k_1)\,I\, u(k'_1)\Big)\,
I^{(\mu\to 0)}_N(q^2_\perp) \Big(\Phi_{\nu}^+(k_2)g_{\nu\nu'}\Phi_{\nu'}
(k'_2)\Big),
\eea
V-exchange:
\bea
&&
\Phi^+_\nu(k_2)\bar u(k_1)\, V_{qD^1_1 ;\nu\nu'}^{(V-ex)}(k_1,k_2;k'_1,k'_2)\,
 u(k'_1)\Phi_{\nu'}(k'_2)
\nn\\
&=&
-\sum\limits_N \beta^{(1N)}_V(q^2_\perp)
\Big(\bar u(k_1)\gamma_\xi
u(k'_1)\Big) I^{(\mu\to 0)}_N(q^2_\perp)
\Big(\Phi_{\nu}^+(k_2)g_{\nu\nu'}\Phi_{\nu'}(k'_2)\Big)(k_2+k'_2)_\xi \nn\\
&\,&-\sum\limits_N \beta^{(2N)}_V(q^2_\perp)
\Big(\bar u(k_1)\gamma_\xi
u(k'_1)\Big) I^{(\mu\to 0)}_N(q^2_\perp)
\Big[\Phi_{\xi}^+(k_2)\Big(k_{2\nu}g_{\nu\nu'}\Phi_{\nu'}(k'_2)\Big)\nn\\
&\,&+
\Big(\Phi_{\nu}^+(k_2)g_{\nu\nu'}k'_{2\nu'}\Big)\Phi_{\xi}(k'_2)
\Big].
\eea
 We can re-write the sum of interaction terms in (\ref{aD1-1-6}) in a
compact form:
\bea \label{aD1-1-7} &\,&\Phi^+(k_2)\bar u(k_1)\,
V_{qD^1_1;\nu\nu'}(k_1,k_2;k'_1,k'_2)\, u(k'_1)\Phi(k'_2),
\nn \\
&&V_{qD^1_1;\nu\nu'}(k_1,k_2;k'_1,k'_2)=
V_{qD^1_1 ;\nu\nu'}^{(S-ex)}(k_1,k_2;k'_1,k'_2)+
V_{qD^1_1 ;\nu\nu'}^{(V-ex)}(k_1,k_2;k'_1,k'_2).\qquad\qquad
\eea

\subsubsection{Spectral integral equation for $qD^1_1$  system   }

The spectral integral equation for $qD^1_1$  system with fixed total
spin $S=1/2,3/2$, orbital momentum $L$ and total angular momentum
 $J^P$ reads:
\bea  \label{aDA-34}
&&
\Phi_\nu^+(k_2)\bar u(k_1)S_{\nu\mu_1\ldots\mu_{j}}^{(S,L,J^P)}(k_1,k_2)
G^{(S,L,J^P)}(k^2_\perp)
\psi_{(J^P)\mu_1\ldots\mu_{j}}(p)
\nn\\
&&=\Phi_\nu^+(k_2)\bar u(k_1)\int\limits^\infty_{(m+M_{D^1_1})^2}
\frac{ds'}{\pi}\frac{d\phi_2(P';k'_1,k'_2)}{s'-M^2_{S,L,J^P}}
 V_{qD^1_1;\nu\nu'}(k_1,k_2;k'_1,k'_2)\frac{\hat k'_1 +m}{2m}
\nn\\
&&\times S_{\nu'\mu'_1\ldots\mu'_{j}}^{(S,L,J^P)}(k'_1,k'_2)G^{(S,L,J^P)}(k'^2_\perp)
\psi_{\mu'_1\ldots\mu'_j}^{(J^P)}(p).
\eea
Here, as in  case of the  $qD^0_0$ system, the interaction block
(the right-hand side of (\ref{aDA-34}))
is written with the use of  spectral integral over $ds'$, and
$d\phi_2(P';k'_1,k'_2)$ is the phase space, see Eq. (\ref{A-34ps}), for
the $qD^1_1$  system in the intermediate state.

As in Eq. (\ref{A-34}), we can transform Eq. (\ref{aDA-34}) convoluting
the left-hand and right-hand sides with spin operator of vertices, see (\ref{aDA-23})
and (\ref{aDA-24}),
and integrating the  convoluted terms over final-state phase space
of the $qD^1_1$  system, $d\phi_2(p;k_1,k_2)$.
We obtain:
\bea  \label{aDA-36}
&&\int d\phi_2(p;k_1,k_2)
Sp\Big[S^{(S,L,J^P)}_{(qD^1_1)\nu\nu_1\ldots\nu_j}(k_1,k_2)
\frac{\hat k_1+m}{2m}S^{(S,L,J^P)}_{(qD^1_1)\nu\mu_1\ldots\mu_j}(k_1,k_2)
F_{\mu_1\ldots\mu_j}^{\nu_1\ldots\nu_j}(p)\Big]\nn \\
&&\times (s-M^2_{S,L,J^P})
\psi^{(L,J^P)}_{(qD^1_1)}(s)
\nn\\
&&=\int d\phi_2(p;k_1,k_2)Sp\Big[
S^{(S,L,J^P)}_{(qD^1_1)\nu\nu_1\ldots\nu_j}(k_1,k_2)\frac{\hat k_1+m}{2m}
\int\limits^\infty_{(m+M_{D^1_1})^2}
\frac{ds'}{\pi}d\phi_2(P';k'_1,k'_2) \nn\\
&&\times V_{qD^1_1;\nu\nu'}(k_1,k_2;k'_1,k'_2)\frac{\hat k'_1 +m}{2m}
S^{(S,L,J^P)}_{(qD^1_1)\nu'\mu'_1\ldots\mu'_j}(k'_1,k'_2)\psi^{(L,J^P)}_{(qD^1_1)}(s')
F_{\mu'_1\ldots\mu'_j}^{\nu_1\ldots\nu_j}(p)
\Big] .\quad\qquad
\eea

\begin{figure}[h]
\centerline{\epsfig{file=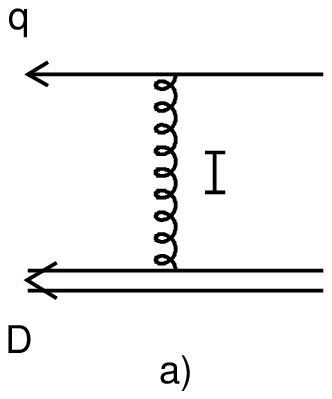,width=4cm}
            \epsfig{file=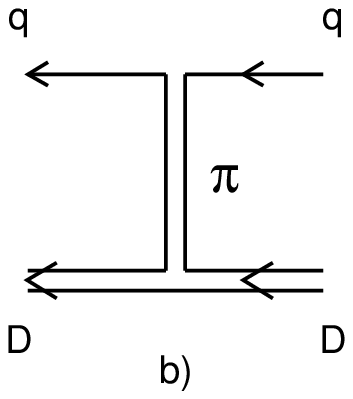,width=4cm}
            \epsfig{file=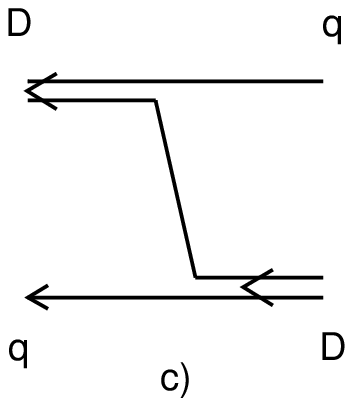,width=4cm}}
\caption{a) Confinement interaction  taken into account
in Eqs. (\ref{A-34}) and (\ref{aDA-34}); b,c) pion-exchange
and quark-exchange interactions which are supposed to be small for highly
excited $qD^0_0$ and  $qD^1_1$  systems. }\label{4} \end{figure}

\section{Conclusion}

We have derived spectral integral equations for the simplest case, when
quark--diquark system is a one-channel system: this is
$q D^0_0$ for nucleon states and $q D^1_1$ for $\Delta_J$ states.

Considering quark--diquark states, we take into account the confinement
interaction only (Fig. \ref{4}a) that is a rather rough
approximation. Still,
the above-performed classification of baryon states \cite{qd,qd2} gives
us a hint that such an approximation may work qualitatively.
More precise results need including and investigating  other
interactions, for example,
the pion and $u$-channel quark exchanges (Figs.  \ref{4}b and \ref{4}c)
 as perturbative admixture -- for more detail see discussion in
 \cite{qd}.

\section*{Acknowledgments}

We thank L.G. Dakhno
for helpful discussions. The paper was supported by the RFFI grants
07-02-01196-a and RSGSS-3628.2008.2.

 \end{document}